\documentclass{epl}
\usepackage{amssymb,amsmath}
\usepackage[T1]{fontenc}

\title{Interplay between incommensurate phases in the cuprates}

\author{ Marcin Raczkowski\inst{1} \and 
         Raymond Fr\'esard\inst{2} \and 
         Andrzej M. Ole\'s\inst{1,3}  }

\shortauthor{M. Raczkowski \etal}
\shorttitle{ Interplay between incommensurate phases in the cuprates }

\institute{                    
  \inst{1} Marian Smoluchowski Institute of Physics, Jagellonian
           University, ul. Reymonta 4, PL-30059 Krak\'ow, Poland \\
  \inst{2} Laboratoire CRISMAT, UMR CNRS--ENSICAEN(ISMRA) 6508,
           6 Bld. du Mar\'echal Juin, F-14050 Caen, France \\
  \inst{3} Max-Planck-Institut f\"ur Festk\"orperforschung,
           Heisenbergstrasse 1, D-70569 Stuttgart, Germany 
}


\begin{document}

\maketitle

\begin{abstract}
We establish the qualitative behavior of the incommensurability 
$\epsilon$, optimal domain wall filling $\nu$ and chemical potential 
$\mu$ for increasing doping by a systematic slave-boson study of an 
array of vertical stripes separated by up to $d=11$ lattice constants. 
Our findings obtained in the Hubbard model with the next-nearest 
neighbor hopping $t'=-0.15t$ agree qualitatively with the experimental 
data for the cuprates in the doping regime $x\lesssim 1/8$. 
It is found that $t'$ modifies the optimal filling $\nu$ and triggers 
the crossover to the diagonal (1,1) spiral phase at increasing doping, 
stabilized already at $x\simeq 0.09$ for $t'=-0.3t$.
\end{abstract}

%%%%%%%%%%%%%%%%%%%%%%%%%%%%%%%%%%%%%%%%%%%%%%%%%%%%%%%%%%%%%%%%%%%%%%%%
%%                           Introduction
%%%%%%%%%%%%%%%%%%%%%%%%%%%%%%%%%%%%%%%%%%%%%%%%%%%%%%%%%%%%%%%%%%%%%%%%
Interest in transition metal oxides has never been restricted to the 
most spectacular phenomenon of the high-$T_c$ superconductivity, but 
also concerns, \emph{inter alia}, metal-insulator transitions, colossal 
magnetoresistance, and orbital ordering \cite{Mae04}. Among these 
phenomena the so-called stripe phases attract much attention. 
In this context the recent indications of universal magnetic excitations 
in doped cuprates are intriguing \cite{Lee06}. Remarkably, some features 
of the magnetic spectra including  their anisotropic two-dimensional 
character established in a detwinned YBa$_{2}$Cu$_3$O$_{6+\delta}$ 
(YBCO) sample \cite{Hin04} can be understood in terms of fluctuating 
stripes suggesting that they are one of generic properties of the copper 
oxides \cite{Sei05}. 

Typically, such states result from the competition between the 
superexchange interaction, which stabilize the antiferromagnetic (AF) 
long-range order in the parent Mott insulator, and the kinetic energy 
of doped holes. Indeed, the magnetic energy is gained when electrons 
occupy the neighboring sites and their spins coupled by the 
superexchange order as in the N\'eel state, whereas the kinetic energy 
is gained when the holes can move and the AF order is locally 
suppressed along a domain wall (DW), leading to the formation of 
\emph{site-centered} (SC) stripes. Other possible structures are 
\emph{bond-centered} (BC) stripes. In this case, AF domains with a 
lower hole density and a stronger spin polarization are separated by 
DWs given by ladders with an increased hole density and a weak 
ferromagnetic order on the rungs.  

The most direct evidence for stripe phases in doped antiferromagnets 
has come from neutron scattering studies in which charge and spin 
modulations are identified by the appearance of several extra 
incommensurate (IC) Bragg peaks \cite{Kiv03}. Indeed, neutron 
diffraction measurements performed on La$_{1.6-x}$Nd$_{0.4}$Sr$_x$CuO$_4$ 
(Nd-LSCO), a model compound for which the evidence of vertical spin and 
charge stripe order is strongest, revealed that magnetic peaks
are displaced from the AF maximum at ${\bf Q}_{\rm AF}=(\pi,\pi)$ 
to the points ${\bf Q}=\pi(1\pm2\epsilon,1)$ and 
${\bf Q}=\pi(1,1\pm2\epsilon)$ \cite{Tra95}. 
Remarkably, the incommensurability $\epsilon$ varies linearly with 
doping $\epsilon=x$ in the underdoped regime of $x<1/8$ meaning 
a fixed stripe filling $\nu=1/2$. A similar value of the number of 
holes per DW $\nu=0.59$ has also been estimated in a recent resonant 
soft X-ray scattering study of La$_{2-x}$Ba$_x$CuO$_4$ (LBCO) which is 
a more direct evidence of charge modulation \cite{Abb05}.  
In contrast, beyond $x=1/8$, one finds in experiment a lock-in effect 
with $\epsilon=1/8$, corresponding to a robust stripe phase with 
a charge (magnetic) unit cell 
consisting of four (eight) sites, and the AF domains with three atoms 
along the $x$ direction. The essentially identical modulation and 
doping dependence of $\epsilon$ was observed in superconducting 
crystals of La$_{2-x}$Sr$_x$CuO$_4$ (LSCO) with $x>0.05$ \cite{Yam98}.
Also in YBCO the incommensurability increases first linearly and next 
saturates at the effective hole doping $x\simeq 0.10$ \cite{Dai01}. 

Apart from the neutron scattering, stripe phases have also measurable 
consequences in angle-resolved photoemission spectroscopy (ARPES) 
\cite{Dam03}. Indeed, the low-energy spectral weight of Nd-LSCO at 
$x=1/8$, is mostly concentrated along the antinodal $\Gamma-X$ and 
$\Gamma-Y$ directions, while there is a distinct gap for charge 
excitations around the $S=(\pi/2,\pi/2)$ point, as expected in the SC 
stripe picture \cite{Wro00,Zac00,Fle00}. In contrast, ARPES spectra of 
both LSCO and Nd-LSCO at $x=0.15$ have revealed the existence of 
appreciable spectral weight along the nodal $\Gamma-S$ direction. 
Therefore, as the BC stripes reproduce quite well the nodal segments 
\cite{Zac00}, it seems that upon increasing doping they are formed at 
the expense of the SC ones. The relevance of the BC stripes at the 
doping level $x=0.15$ is also supported by recent studies which have 
yielded pronounced spectral weight both in the nodal and antinodal 
directions, reproducing quite well the experimental results in Nd-LSCO 
and LSCO \cite{Wro06}. 

An alternative scenario which might explain the IC spin structure is 
a deformation of the AF order which optimizes a hole motion within   
a spiral phase \cite{Shr89}. The interest in the spiral state was 
recently renewed due to experimental results indicating a spin glass 
behavior of LSCO at small doping consistent with this scenario 
\cite{Has04}. On the one hand, observation of the charge order in 
Nd-LSCO and LBCO rules out the spiral phase since the density of holes 
is then expected to be uniform. On the other hand, the question whether 
the charge order is a generic feature of the cuprates is far from being 
resolved yet \cite{Kiv03}. In fact, Lindg\aa rd \cite{Lin05} 
has shown that spiral states can also resolve the universality of 
magnetic excitations in the cuprates and provide a competing paradigm 
with the stripe phase concept. 

Hence, an interplay between domain wall structures and spiral phases 
poses an interesting problem to study \cite{Gia90} even though the 
searching for the optimal filling of domain walls corresponding to the 
true ground state is rather a formidable task. Therefore, we limit 
ourselves to vertical site-centered (VSC) and vertical 
bond-centered (VBC) stripe phases, with the size of the AF domains 
varying from $d=11$ to $3$ lattice constants. In both structures, the 
largest distance $d=11$ corresponds to a unit cell with $22$ atoms. Note 
that the same length of the magnetic unit cell for a fixed $d$ makes the 
SC and BC structures practically indistinguishable from each other
in neutron diffraction experiments.

%%%%%%%%%%%%%%%%%%%%%%%%%%%%%%%%%%%%%%%%%%%%%%%%%%%%%%%%%%%%%%%%%%%%%%%%
%%                            The model            
%%%%%%%%%%%%%%%%%%%%%%%%%%%%%%%%%%%%%%%%%%%%%%%%%%%%%%%%%%%%%%%%%%%%%%%%
We study the stripe and spiral phases using the Hubbard model, 
\begin{equation}
H=-t\sum_{\langle ij\rangle\sigma}
         (c^{\dag}_{i\sigma}c^{}_{j\sigma}+H.c.)
-t'\sum_{\langle\langle ij\rangle\rangle\sigma}	 
         (c^{\dag}_{i\sigma}c^{}_{j\sigma}+H.c.)
+U\sum_{i}n^{}_{i\uparrow}n^{}_{i\downarrow},
\label{eq:Hubb0}
\end{equation}
where the electron hopping $t$ involves the nearest neighbor sites 
$\langle ij\rangle$, $t'$ the next-nearest neighbor sites 
$\langle\langle ij\rangle\rangle$, and $U$ stands for the 
on-site Coulomb interaction. 
There are several experimental and theoretical studies suggesting that 
$t'$ is finite in the cuprates. For example, topology of the Fermi 
surface seen by ARPES \cite{Dam03} can be only understood by introducing 
$t'$ \cite{Toh04}. It also offers an explanation for the variation 
of $T_c$ among different families of hole-doped cuprates \cite{Pre05}. 
Moreover, exact diagonalization (ED) studies have shown that while the 
$d$-wave superconductivity correlation is slightly suppressed 
by $t'<0$ in the underdoped region, it is substantially enhanced 
in the optimally doped and overdoped regions, indicating that
$t'$ is of great importance for the pairing instability \cite{Shi04}.

We employ a rotationally invariant version of the slave-boson (SB) 
approach in spin space \cite{Fre92}, which opens a possibility of 
studying both noncanted stripe phases and spiral order on equal footing. 
More details on the calculation method can be found in ref. 
\cite{Zim97}. Calculations were carried out on square clusters with the 
linear dimension along the $x$ direction (perpendicular to the DWs) 
chosen as an \emph{even} multiplicity of the elementary stripe unit cell 
dimension in all cases, so that the cluster size used for the considered 
unit cells varied from 128$\times$128 up to 144$\times$144. 
These calculations became possible by developing 
an efficient scheme in the reciprocal space which makes use of the 
stripe symmetry \cite{Rac06}. For so large systems the finite size 
effects are below 10$^{-5}t$ and one is able to obtain a realistic 
comparison of the free energies of structures with different size of 
the unit cell. In the numerical studies, we have chosen $U=12t$, which 
gives the ratio of $J/t=1/3$ (with the superexchange $J=4t^2/U$), 
being a value representative for LSCO \cite{Jef92}.

%%%%%%%%%%%%%%%%%%%%%%%%%%%%%%%%%%%%%%%%%%%%%%%%%%%%%%%%%%%%%%%%%%%%%%%%
%%                           Free energy gain              
%%%%%%%%%%%%%%%%%%%%%%%%%%%%%%%%%%%%%%%%%%%%%%%%%%%%%%%%%%%%%%%%%%%%%%%%

%%%%%%%%%%%%%%%%%%%%%%%%%%%%%%%%%%%%%%%%%%%%%%%%%%%%%%%%%%%%%%%%%%%%%%%%
%%                             figure 1 & 2
%%%%%%%%%%%%%%%%%%%%%%%%%%%%%%%%%%%%%%%%%%%%%%%%%%%%%%%%%%%%%%%%%%%%%%%%
\begin{figure}
\twofigures[scale=0.38]{fig1.eps}{fig2.eps}
\caption
{(Color online) 
Free energy gain $\delta F$ per site in the VSC stripe phases with 
respect to the AF phase as a function of doping $x$, as obtained for the 
$t$-$t'$-$U$ model (1) with $U=12t$ and: (a) $t'=0$; (b) $t'=-0.15t$. 
Domain walls are separated by $d=3,\dotsc, 11$ lattice constants. 
Circles and squares show the corresponding data for (1,0) and (1,1) 
spiral order, respectively. 
}
\label{fig:fig1}
\caption
{(Color online) 
Free energy gain $\delta F$ per site as in fig.~\ref{fig:fig1}, obtained 
for the $t$-$t'$-$U$ model (1) with $U=12t$ and $t'=-0.3t$ for: 
(a) VSC stripe phases; (b) VBC stripe phases. 
}
\label{fig:fig2}
\end{figure}

We begin with the free energy gain of the VSC stripe phase with respect 
to the AF phase, $\delta F\equiv F - F_{\rm AF}$, shown in fig. 
\ref{fig:fig1} as a function of doping $x$ for representative values 
of the next-nearest neighbor hopping, i.e., $t'=0$, $t'=-0.15t$ 
(VBC phases are addressed below). Quite generally, one observes that 
the energy favors the stripe phases with the largest distance $d=11$ 
between DWs for the lowest values of $x$, and the curves for next 
smaller $d-1$ systematically cross the ones for $d$ upon increasing 
doping, meaning that stripe phases with gradually decreasing AF domains 
become optimal phases. 
For $t'=0$, this effect continues until the $d=4$ stripe phase is 
reached as the $d=3$ stripe phase is a highly-excited state in this 
case. On the contrary, finite $t'$ results in a deeper energy minimum 
of the $d=3$ phase and stabilizes it in the overdoped regime $x\ge 0.2$. 
Next, except for the smallest $d\leq 4$ cases, increasing $|t'|$ shifts 
the free energy minima towards lower doping level which should affect 
the filling of the DWs and the charge distribution in the stripe ground
state. Remarkably, for $t'=-0.15t$, a value very close to that derived 
for LSCO \cite{Pav01}, one finds that the most stable stripes 
are separated by $d=4$ lattice spacings in a sizeable doping range 
above $x\simeq 1/8$, both for the SC and BC stripe phase, in agreement 
with the neutron scattering experiment \cite{Yam98} and with dynamical 
mean field theory (DMFT) calculations \cite{Fle00} for LSCO.  

Regarding the spiral phases, upon hole doping away from half-filling the 
AF order becomes immediately unstable towards a diagonal (1,1) spiral 
order with ${\bf Q}=\pi(1-2\epsilon,1-2\epsilon)$ and then at higher 
doping $x\simeq 0.18$ towards a vertical (1,0) one characterized by 
${\bf Q}=\pi(1-2\epsilon,1)$, as suggested by early SB studies of the 
Hubbard model \cite{Fre91}. However, a finite next-nearest neighbor 
$t'=-0.15t$ has severe consequences for the interplay between both 
spiral phases and shifts the crossover towards a higher doping 
$x\simeq 0.23$ in agreement with the SB studies of the $t$-$t'$-$J$ 
model \cite{Dee94}. It also affects the competition between the spiral 
and stripe phases so that at $x>1/8$ the latter become unstable with 
respect to the (1,1) spirals. However, the energy difference between the 
most stable stripe phase with $d=4$ and the lowest energy spiral phase 
is less than $0.002t$, which is considered to be too small to lead to
an unambiguous conclusion. Indeed, comparison of the SB data with 
available ED results on a $4\times 4$ cluster obtained in the Hubbard 
model at $t'=0$ and $x=1/8$ doping \cite{Fan90} shows that taking into 
account stripe phases allows one to approach closer the ED energy than 
the best spiral phase but due to the quantum fluctuations the energy 
difference remains at the 9\% level (\textit{cf}. table~\ref{tab:Ftp0}). 

%%%%%%%%%%%%%%%%%%%%%%%%%%%%%%%%%%%%%%%%%%%%%%%%%%%%%%%%%%%%%%%%%%%%%%%%
%%                             table 1
%%%%%%%%%%%%%%%%%%%%%%%%%%%%%%%%%%%%%%%%%%%%%%%%%%%%%%%%%%%%%%%%%%%%%%%%
\begin{table}[b!]
\caption 
{
Comparison of the free energy $F$ per site for various phases as found 
within the SB approximation in the Hubbard model at $x=1/8$ with the 
ED data of ref. \cite{Fan90}. The ground state 
(VBC and VSC) stripe phases are separated by $d=7$ lattice constants. 
Parameters: $U=12t$, $t'=0$.
}
\label{tab:Ftp0}
\begin{center}
\begin{tabular}{ccccccccc}
\hline
phase &&     AF   &   (1,0)  &  (1,1)   &    VBC   &  VSC     && ED [29] \\
\hline
$F/t$ && $-$0.5393&$-$0.5613 &$-$0.5700 &$-$0.5756 &$-$0.5756 &&$-$0.6282\\
\hline
\end{tabular}
\end{center}
\end{table}

%%%%%%%%%%%%%%%%%%%%%%%%%%%%%%%%%%%%%%%%%%%%%%%%%%%%%%%%%%%%%%%%%%%%%%%%
%%                             table 2
%%%%%%%%%%%%%%%%%%%%%%%%%%%%%%%%%%%%%%%%%%%%%%%%%%%%%%%%%%%%%%%%%%%%%%%%
\begin{table}[b!]
\caption 
{
Comparison of the ground state free energy $F$ per site for the VSC and 
VBC stripe phases as found in the $t$-$t'$-$U$ model with $U=12t$ and: 
$t'=-0.15t$ and $t'=-0.3t$. 
}
\label{tab:Ftps}
\begin{center}
\begin{tabular}{cccccccccccc}
\hline
\multicolumn{2}{c}{}              &
\multicolumn{4}{c}{$t'=-0.15t$}   &
\multicolumn{1}{c}{}              &
\multicolumn{4}{c}{$t'=-0.3t$}    \\
\multicolumn{2}{c}{}              &
\multicolumn{2}{c}{VSC}           &
\multicolumn{2}{c}{VBC}           &
\multicolumn{1}{c}{}              &
\multicolumn{2}{c}{VSC}           &
\multicolumn{2}{c}{VBC}           \\
\hline
\multicolumn{1}{c}{$x$}           &
\multicolumn{1}{c}{}              &
\multicolumn{1}{c}{$d$}           &
\multicolumn{1}{c}{$F/t$}         &
\multicolumn{1}{c}{$d$}           &
\multicolumn{1}{c}{$F/t$}         &
\multicolumn{1}{c}{}              &
\multicolumn{1}{c}{$d$}           &
\multicolumn{1}{c}{$F/t$}         &
\multicolumn{1}{c}{$d$}           &
\multicolumn{1}{c}{$F/t$}         \\ 
\hline
0.050 && 11 & $-$0.4263 & 11 & $-$0.4263 &&  9 & $-$0.4289 & 10 & $-$0.4280 \\ 
0.055 && 10 & $-$0.4360 & 10 & $-$0.4359 &&  8 & $-$0.4388 &  9 & $-$0.4378 \\
0.060 &&  9 & $-$0.4456 &  9 & $-$0.4455 &&  7 & $-$0.4488 &  8 & $-$0.4477 \\
0.070 &&  8 & $-$0.4649 &  8 & $-$0.4648 &&  6 & $-$0.4686 &  7 & $-$0.4673 \\
0.080 &&  7 & $-$0.4841 &  7 & $-$0.4840 &&  5 & $-$0.4882 &  6 & $-$0.4869 \\
0.090 &&  6 & $-$0.5034 &  6 & $-$0.5032 &&  5 & $-$0.5080 &  5 & $-$0.5064 \\
0.100 &&  5 & $-$0.5225 &  6 & $-$0.5224 &&  4 & $-$0.5275 &  5 & $-$0.5260 \\
0.120 &&  5 & $-$0.5607 &  5 & $-$0.5607 &&  4 & $-$0.5659 &  4 & $-$0.5645 \\
0.140 &&  4 & $-$0.5985 &  4 & $-$0.5983 &&  3 & $-$0.6022 &  4 & $-$0.6005 \\
0.160 &&  4 & $-$0.6342 &  4 & $-$0.6342 &&  3 & $-$0.6370 &  3 & $-$0.6344 \\
0.180 &&  4 & $-$0.6671 &  4 & $-$0.6670 &&  3 & $-$0.6678 &  3 & $-$0.6672 \\
0.200 &&  3 & $-$0.6978 &  3 & $-$0.6983 &&  3 & $-$0.6949 &  3 & $-$0.6956 \\
0.250 &&  3 & $-$0.7682 &  3 & $-$0.7689 &&  3 & $-$0.7475 &  3 & $-$0.7493 \\
0.300 &&  3 & $-$0.8242 &  3 & $-$0.8245 &&  3 & $-$0.7822 &  3 & $-$0.7833 \\ 
\hline
\end{tabular}
\end{center}
\end{table}

In contrast, a larger value of the ratio $|t'/t|=0.3$ as expected for 
YBCO \cite{Pav01}, clearly drives the system towards the diagonal (1,1) 
spiral phase and the crossover from the VSC (VBC) stripe phase appears 
already slightly below (above) $x=0.09$, respectively, as depicted in 
fig.~\ref{fig:fig2}. 
Note that in the case of the SC DWs it is the $d=4$ stripe phase which 
is unstable towards the spiral order and indeed the smallest distance 
between DWs established in YBCO corresponds to $d=5$ \cite{Dai01}.
Our results are then consistent with both density matrix renormalization 
group \cite{Whi99} and Hartree-Fock \cite{Nor01} studies indicating the
suppression of the stripe phases with increasing $|t'|$. Moreover, 
large $|t'|$ removes also the degeneracy between the two stripe 
structures which shows up at $t'=-0.15t$. Indeed, as reported in 
table~\ref{tab:Ftps} the energy difference between the lowest energy 
SC and BC configuration in the doping regime $x<0.2$ is comparable to 
the accuracy of the present calculations. In contrast, when $t'=-0.3t$, 
the SC stripe phase is noticeably more stable than its BC counterpart 
even for $x<0.2$. The crossover towards the BC stripes above $x=0.2$ 
remains in agreement with a recent analysis of the propagation of a hole 
inside the BC DW which has shown that such a DW structure enables 
a larger kinetic energy gain than a narrower SC one, which becomes 
especially important at a large doping level when the distances between 
stripes are small \cite{Wro06a}. 

%%%%%%%%%%%%%%%%%%%%%%%%%%%%%%%%%%%%%%%%%%%%%%%%%%%%%%%%%%%%%%%%%%%%%%%%
%            Properties of the stripe ground state                  
%%%%%%%%%%%%%%%%%%%%%%%%%%%%%%%%%%%%%%%%%%%%%%%%%%%%%%%%%%%%%%%%%%%%%%%%

%%%%%%%%%%%%%%%%%%%%%%%%%%%%%%%%%%%%%%%%%%%%%%%%%%%%%%%%%%%%%%%%%%%%%%%%
%%                             figure 3 & 4
%%%%%%%%%%%%%%%%%%%%%%%%%%%%%%%%%%%%%%%%%%%%%%%%%%%%%%%%%%%%%%%%%%%%%%%%
\begin{figure}
\twofigures[scale=0.47]{fig3.eps}{fig4.eps}
\caption
{
(Color online) 
Doping dependence of: (a,b) the magnetic incommensurability $\epsilon$, 
and (c,d) stripe filling $\nu$ for the VSC (left) as well as VBC (right) 
stripe ground state deduced from figs. \ref{fig:fig1} and \ref{fig:fig2} 
for $t'=0$ (circles), $t'=-0.15t$ (squares), and $t'=-0.3t$ (triangles). 
Solid line in (a,b) shows the idealized experimental behavior of 
$\epsilon$ in LSCO \cite{Yam98}.
}
\label{fig:fig3}
\caption
{
(Color online) 
Doping dependence of the magnetic incommensurability $\epsilon$ as 
obtained at $U=12t$ for: (a) (1,0) and (b) (1,1) spiral phase. 
Meaning of symbols as in fig.~\ref{fig:fig3}. 
}
\label{fig:fig4}
\end{figure}

The experimental data should be compared with our findings concerning 
the doping dependence of the magnetic incommensurability $\epsilon=1/2d$ 
and the optimal stripe filling, $\nu=N_{\rm h}/(N_{y} N_{\rm DW})$,
where $N_{\rm h}$ is the hole excess compared to half-filling, whereas 
$N_y$ stands for the actual length of the cluster with $N_{\rm DW}$ 
DWs along the $y$ direction. The ground state properties of both 
SC and BC phases are shown in fig.~\ref{fig:fig3}. The points in 
fig. \ref{fig:fig3} were deduced from figs.~\ref{fig:fig1} and 
\ref{fig:fig2} and correspond to the middle of the stability region of 
the lowest energy configuration. The only exception is the $d=3$ case 
with $t'=-0.3t$ in which they are plotted for the minimum of the free 
energy. Such a choice guarantees that, at each particular doping level, 
a considered stripe phase with a given periodicity $d$ would be indeed 
realized at least in the vast majority of the system. 
Comparing fig.~\ref{fig:fig3} with the experimental results in LSCO 
\cite{Abb05,Yam98}, one finds that the robust stability of the 
half-filled ($\nu=1/2$) $d=4$ stripes at $x=1/8$ requires, in agreement 
with the previous SB studies on small 16$\times$16 clusters \cite{Sei04}, 
a finite next-nearest neighbor hopping $t'<-0.15t$. 
In fact, for $t'=-0.15t$ present calculations give almost a linear 
dependence $\epsilon=x$ for $x\lesssim 1/8$ and may be considered 
as reproducing a lock-in effect in a sizeable doping range above 
$x\simeq 1/8$ until the $d=3$ stripe phase sets in. 
We emphasize that stable $d=3$ stripe phases with $\epsilon=1/6$ have 
also been found in the SB studies of the three-band Hubbard model in 
the doping regime $x>0.225$ \cite{Lor02}.
The apparent absence of this phase in the experimental data for LSCO
could follow from two effects which go beyond the present study: 
 ($i$) repulsion between DWs, and 
($ii$) quantum fluctuations which destabilize the ladder-like AF 
       domains.       

Remarkably, in the regime where $\epsilon$ follows linearly $x$, 
an increasing density of stripes allows the system to maintain fixed 
filling $\nu$, but its actual value strongly depends on $t'$. This, 
in turn, results in almost doping independent chemical potential $\mu$ 
explaining the experimentally observed pinning of $\mu$ \cite{Ino97}. 
In contrast, in the overdoped region with a lock-in effect of $\epsilon$, 
the size of the AF domains saturates, doped holes penetrate into the AF 
domains, and $\mu$ varies fast with doping. Unfortunately, for 
$t'=-0.15t$, the established shift of $\mu$ exceeds the experimental 
value by a factor close to 2. Therefore, the present effective model can 
only explain qualitative trends and one needs to carry out calculations 
within more realistic multiband models including oxygen orbitals 
in order to reproduce quantitatively the experimental data.

Finite $t'<0$ also modifies the spiral order --- here one finds that 
the deviation of the spiral wavevector ${\bf Q}$ from the AF one strongly 
accelerates, as shown in fig.~\ref{fig:fig4}. It is worth noticing 
that the increase of the incommensurability with increasing $|t'|$ is 
consistent both with quantum Monte Carlo \cite{Duf95} and with DMFT 
\cite{Fle99} results.

Summarizing, we have performed systematic studies of incommensurate 
phases by considering variable size of AF domains in stripe phases, 
as well as the spiral order. Our findings obtained in the Hubbard model 
within the SB approach for $t'=-0.15t$ agree qualitatively in the low 
doping regime $x\lesssim 1/8$ with the experimental data for the cuprates 
and reveal a strong influence of the next-nearest neighbor hopping $t'$ 
on the optimal filling of DWs. Simultaneously, finite $t'<0$ promotes
the onset of the diagonal spiral phases and increases the 
optimal spiral pitch $\epsilon$. We therefore conclude that a large 
value of $|t'|$ might be the reason why the static charge order has been 
detected in YBCO only in the highly underdoped regime \cite{Dai02}.   

\acknowledgments

MR acknowledges the hospitality of the Laboratoire CRISMAT where part of 
this work has been done. This work was supported by the Polish Ministry 
of Science and Education under Project No. 1~P03B~068~26, and by the 
Minist\`ere Fran\c{c}ais des Affaires Etrang\`eres under POLONIUM 09294VH.

\end{document}